\documentclass[10pt,journal,compsoc]{IEEEtran}
\usepackage[T1]{fontenc}

\ifCLASSINFOpdf
\else
\fi
\usepackage{graphicx}
\usepackage{amsmath}
\usepackage{subfigure}
\usepackage{multirow}
\usepackage{graphicx}
\usepackage{amsmath}
\usepackage{multirow}
\usepackage{amsfonts,amsmath}
\usepackage{amssymb, nccmath}
\DeclareMathAlphabet\mathchorus     {T1}{qzc} {m} {n}
\usepackage{amssymb, nccmath}
\usepackage{dsfont}
\usepackage{makecell}
\usepackage{booktabs}
\usepackage[utf8]{inputenc}
\usepackage{dsfont}
\usepackage{booktabs}
\usepackage{graphicx}
\usepackage{grffile}
\usepackage{boldline}
\usepackage{grffile}
\usepackage{amsthm}
\newtheorem{definition}{Definition}
\usepackage{array} 
\usepackage{mathtools}
\usepackage{fontenc}
\usepackage{tikz}
\usepackage{supertabular,enumitem}
\usepackage{tikz}

\usepackage{pgfplots}

\definecolor{mygreen}{HTML}{03C03C}
\definecolor{mypink}{HTML}{008000}
\definecolor{Mycolor}{HTML}{FFC40C}
\usepackage[linesnumbered, ruled]{algorithm2e}

\usepackage{subfig}
\makeatletter
\newcommand{\removelatexerror}{\let\@latex@error\@gobble}
\def\ps@IEEEtitlepagestyle{%
	\def\@oddfoot{\mycopyrightnotice}%
	\def\@oddhead{\hbox{}\@IEEEheaderstyle\leftmark\hfil\thepage}\relax
	\def\@evenhead{\@IEEEheaderstyle\thepage\hfil\leftmark\hbox{}}\relax
	\def\@evenfoot{}%
}

\def\mycopyrightnotice{%
	\begin{minipage}{\textwidth}
		\centering \scriptsize
		This article has been accepted in IEEE Transactions on Services Computing Journal © 2022 IEEE. Personal use of this material is permitted. Permission from
		IEEE must be obtained for all other uses, in any current or future media, including reprinting/republishing this material for advertising or promotional purposes, creating new collective works, for resale or redistribution to servers or lists, or reuse of any copyrighted component of this work in other works. This work is freely available for survey and citation.
		
	\end{minipage}
}
\makeatother
\renewcommand\thesection{\arabic{section}}

\renewcommand{\thesubsubsection}{\Alph{subsubsection}}

\usepackage[cmintegrals]{newtxmath}
\begin{document}
%
\title{A High Availability Management Model based on VM Significance Ranking and Resource Estimation for Cloud Applications}
%
%
%
\author{Deepika~Saxena and
	Ashutosh~Kumar~Singh,~\IEEEmembership{Senior~Member~IEEE}
	\thanks{D. Saxena and A.K. Singh are with the Department of Computer Applications, National Institute of Technology, Kurukshetra, India. E-mail: 13deepikasaxena@gmail.com and ashutosh@nitkkr.ac.in }}

\markboth{IEEE TRANSACTIONS ON SERVICES COMPUTING}{Shell \MakeLowercase{\textit{et al.}}: Bare Demo of IEEEtran.cls for Computer Society Journals}
%





\IEEEtitleabstractindextext{
\begin{abstract}
	Massive upsurge in cloud resource usage stave off service availability resulting into outages, resource contention, and excessive power-consumption. The existing approaches have addressed this challenge by providing multi-cloud, VM migration, and running multiple replicas of each VM which accounts for high expenses of cloud data centre (CDC). In this context, a novel VM Significance Ranking and Resource Estimation based High Availability Management (SRE-HM) Model is proposed to enhance service availability for users with optimized cost for CDC. The model estimates 
	resource contention based server failure and organises needed resources beforehand for maintaining desired level of service availability. A significance ranking parameter is introduced and computed  for each VM, executing critical or non-critical tasks followed by the selection of  
	an admissible High Availability (HA) strategy respective to its significance and user specified constraints. It enables cost optimization for CDC by rendering failure tolerance strategies for significant VMs only instead of all the VMs. The proposed model is evaluated and compared against state-of-the-arts by executing experiments using Google Cluster dataset. SRE-HM improved the services availability up to 19.56\% and scales down the number of active servers and power-consumption up to 26.67\% and 19.1\%, respectively over HA without SRE-HM. 
	
\end{abstract}

\begin{IEEEkeywords}
Cloud computing, Cost optimization,  Failure predictor, High Availability,  Resource management, VM Ranking.
\end{IEEEkeywords}}

\maketitle

\IEEEdisplaynontitleabstractindextext
%
\IEEEpeerreviewmaketitle

\section{Introduction}
\IEEEPARstart{N}{owadays}, there is a strong trend towards "digitization in all data and all data in digitization" across the globe, wherein cloud data centres (CDCs) have emerged as a backbone of the entire information and communication technology. It facilitates high availability of information technology (IT) resources and maximum computing benefits at minimum capital investment to the end users. A Cloud Service Providers (CSPs)  is obliged to Service Level Agreement (SLA) adherence by taking responsibility for their infrastructure and ensuring availability and safety at all ends \cite{saxena2020security}, \cite{saxena2022intelligent}, services and performance outage occurs, stemming from a surge in resource utilization \cite{saxena2021op}, \cite{fang2019dsp}, \cite{zhang2015charm}.  However, the exponential growth of internet traffic and cloud resource demands impede the delivery of cloud services with consistent high availability the whole time. 
For instance, a recent survey of the COVID-19 pandemic witnessed numerous spectacular incidents of cloud outages for several hours which have overwhelmed the titan Cloud Service Providers (CSPs), including Zoom, Microsoft Azure, Google Cloud Platform, Amazon Web Services, Salesforce, IBM Cloud, etc. \cite{crn2020outage}. Such incidents prompt a critical question about cloud technology and increase the obligation of high availablity (HA) on CSPs.  Major causes of cloud outages are failures of applications, VMs, and servers or physical machines (PMs) which renders computing and storage instances inaccessible \cite{wang2014festal}, \cite{singh2021quantum}. Such instances of cloud outages point to a critical question on cloud technology concerning reliability and availability of services  for the end users.
 The existing approaches furnish highly available services by applying proactive as well as reactive approaches \cite{mukwevho2018toward}, \cite{endo2016high}, \cite{gill2018failure}. Proactive techniques of failure handling rely on the prior knowledge of the failure of applications and VMs and migrating such VMs. While the reactive techniques include checkpointing, replication, retry, application resubmission, etc. which are triggered at the occurrence of actual failure detection \cite{costa2014system}. Among the proactive approaches, the most reliable approach for high availability is to maintain multiple replicas of VM instances while engaging active physical machines in excess \cite{li2017novel} and hiring multiple clouds \cite{zhang2015charm}. The major concern is that previous reactive approaches engage delay in service provisioning while the proactive approaches account for high operational cost. 
 
\par Since the availability of the cloud resources is measured in terms of mean time between failures ({MTBF}) and mean time to repair (MTTR) associated with service uptime and downtime, respectively \cite{endo2017highly}, \cite{saxena2022ofp}; the key solution for improving services  availability is to maximize MTBF and minimize MTTR i.e., time of recovery/repair by employing fault tolerance strategies along with effective resource management. The proactive resource failure detection and mitigation of VM and server failure enhances service uptime leading to higher MTBF. Diversely, the MTTR can be minimized by appointing multiple replicas for VM instances.  Therefore, both proactive, as well as reactive mechanisms must be applied efficiently to upgrade cloud services availability. 
  A cloud application being composed of numerous small sub-units or tasks makes it highly expensive to provide redundant or alternative instances for all VMs \cite{saxena2021op}, \cite{saxena2021secure}, \cite{zheng2011component}. Moreover, it is not desirable to deliver highly available services for non-critical tasks which either have minimal or no effect on  application execution. In this context, the best approach is to identify the significance of different tasks beforehand and employ a selected optimal HA strategy for their execution  accordingly.  However, this approach is entangled with two primary challenges: (\textit{i}) How to locate the importance of various tasks in a complex cloud application?  (\textit{ii}) How to decide on an admissible HA strategy for different tasks subject to their contribution to an application execution and user-specified constraints?

\subsection{ Our Contribution} 
Contrary to the existing works, this paper proposes a novel  \textbf{S}ignificance \textbf{R}anking and Resource \textbf{E}stimation based \textbf{H}igh {A}vailability \textbf{M}anagement (\textbf{SRE-HM}) model that encapsulates both proactive and reactive approaches to deliver highly available cloud services with optimized operational costs. It locates critical and non-critical tasks  based on their characteristics, number and frequency of invocations. Accordingly, a significance rank computation method is developed for estimation of the importance of different VMs depending on the characteristics of tasks assigned for execution. Thereafter, an optimal HA strategy selection method is established corresponding to significance rank of different VMs and user specified constraints to optimize the operational cost for CSPs.  Further, the proposed model imparts a prior estimation of resource contention and mitigation of performance degradation by migrating VMs and load balancing in real-time. The key contributions of the paper are fourfold:

\begin{itemize}
	\item A Virtual High Availability Zone (VHAZ) comprising of several Virtual High Availability Network (VHAN) is created where a new parameter  `\textit{significance rank}' of a VM is enumerated to locate importance of each VM in an application execution. 
	
	\item A method for selection  of an admissible HA strategy  is developed to curtail the excessive power expenses while achieving high performance and service availability experience for the end users.  
	
	\item A real-time server failure predictor based on Long Short Term Memory (LSTM) Neural Network is implemented to estimate resource-contention and mitigate its worse effects by triggering VM shifting, proactively.

	\item Empirical simulation and performance evaluation of proposed model reveal that it substantially improves high availability for cloud users with optimized cost for cloud  data centre. 

\end{itemize} 
\textit{Organization}: Section 2 discusses related work with research gaps and Section 3 entails description of SRE-HM model. Sections 4, 5, and 6 discuss PM failure estimation and handling, Significance rank computation, and HA strategy selection, respectively. Section 7 presents operational design and complexity analysis followed by the performance evaluation and comparison in Section 8. Section 9 remarks the conclusion and future scope of the proposed model. %
 
 \section{Related work}
 The related work pertaining to HA in cloud environment is distinguished into following sub-sections: (\textit{i}) \textit{Proactive approaches} and (\textit{ii}) \textit{Reactive approaches}.
 \subsection{Proactive approaches} 
 Marahatta et al. \cite{marahatta2020pefs} proposed a deep neural network based failure predictor to differentiate the future tasks on VMs into failure prone and non-failure prone tasks. Accordingly, resources are allocated exclusively for failure and non-failure prone tasks by replicating three consecutive failure-prone tasks into three copies to be executed on different servers to prohibit overlapping and redundant execution. The performance of the work is investigated using Internet and Euler datasets which improve fault tolerance and energy efficiency within CDC.  Pinto et al. \cite{pinto2016hadoop} have developed a Support Vector Machine (SVM) based failure predictor for distributed computing Hadoop clusters which is trained with a non-anomalous dataset during different operation patterns like boot-up, shutdown, idle, task allocation, resource distribution, etc. to detect and classify between normal and abnormal situation. An online prediction based  multi-objective load-balancing (OP-MLB) framework is proposed in \cite{saxena2021op} for proactive estimation of server overload and its alleviation by employing VM migration for improving service availability and energy-efficiency of CDCs. The forthcoming load on VMs is estimated by employing an online  neural-network based prediction system to determine the future resource utilization of the servers proactively. It helped to detect  server overloading  which is tackled by migrating VMs of highest resource capacity from overloaded server to an appropriate server machine. The VM placement and migration are executed using a multi-objective  algorithm  for minimization of power consumption.
 
  \par Sharma et al. presented a failure aware and energy-efficient (FAEE) VM placement scheme in \cite{sharma2019failure} which predicted VM failure using an exponential smoothing based forecasting technique. Thereafter, two fault tolerance methods including VM migration and checkpointing are triggered to handle service failure and upgrade service availability. This work was evaluated using Grid5000 Failure Traffic Analysis (FTA) dataset. It concluded that the energy efficiency and reliability of cloud systems can be significantly improved by considering the failure characteristics of physical resources. 
  Dabbagh et al. \cite{dabbagh2015energy} presented an integrated energy-efficient VM placement and migration framework for the cloud data centre. It applied a Wiener filter based predictor for estimation of the number of VM requests and the future resource requirement. These predicted values are used to allow only the required number of physical machines in active state and helps in achieving a substantial energy saving and resource utilization. Nguyen et al. \cite{hieu2017virtual} addressed service availability problem by adopting multiple usage prediction through  multiple linear regression for estimation of the relationship between the input variables and the output for energy efficient data centres. This work estimated overload or host failure with multiple usage prediction (OHD-MUP) and underloaded host detection with multiple usage prediction (UHD-MUP) and balanced load by migrating selected VMs from overloaded servers to energy-efficient server. 
 Multiple Additive Regression Trees with Gradient Boosting (MART-GB) algorithm is utilized for Cloud Disk Error Forecasting (CDEF) in \cite{xu2018improving} which ranked all disks according to degree of error-proneness to allow shifting
 of existing VMs in real-time and assignment of new VMs on healthy disks to improve service availability.
 Adamu et al. \cite{adamu2017approach} have improved system availability by proposing a Linear Regression (LR) Model and 
 Support Vector Machine (SVM) with a Linear Gaussian kernel for 
 predicting hardware failures in a real-time cloud environment. Sharma et al. \cite{sharma2019failure} have predicted VM failure using a forecasting technique based on exponential smoothing.  This work has utilized two fault tolerance methods including VM migration and checkpointing to handle VM failure and improve service availability. It achieved a significant improvement in energy efficiency and reliability  by considering failure characteristics of physical resources.  A  Fault Tolerant Elastic Resource Management (FT-ERM) framework has been proposed in \cite{saxena2022fault} to improve availability of services by inducing  fault management in servers and VMs. This framework included an online failure predictor for proactive estimation of failure-prone VMs where the operational status of server is monitored with the help of a power analyser, resource estimator, and thermal analyser. The failure-prone VMs are assigned to fault-tolerance unit consisting of decision matrix and safe box to handle any outage beforehand while assuring high availability of services for the cloud users.

 \subsection{Reactive Approaches}
  Zhang et al. \cite{zhang2015charm} proposed a data hosting scheme called as CHARM which collaborated two desired key functions to furnish HA. Firstly, multiple suitable clouds  and an optimal replication scheme for data storage and enhance availability are selected. Secondly, VM migration and re-allocation is triggered as per the variations in data access patterns and pricing of cloud applications.  This work enabled  fine-grained decisions making regarding  which storage mode to be used and which clouds to be used for data placement.
  An HA-aware scheduling algorithm named CHASE is proposed in \cite{jammal2015chase}  that considers and analyses functionality requirements, redundancies between applications, and  real-time interdependencies in real-time. Its prototype was designed to schedule components in real cloud environment
 while communicating with OpenStack. 
 
 \par  Wang et al. \cite{wang2014festal} developed a fault-tolerant elastic scheduling algorithms (FESTAL) to provide a fault tolerant VM scheduling accompanied with virtualization technology and an appropriate VM migration scheme with battery back-up features. It addressed  the issues of
 reliability, elasticity and schedulability of virtualized clouds. It achieved both fault tolerance and
 high performance in terms of resource utilization by employing an elastic resource
 management and the fault-tolerant scheduling algorithms. Extensive
 experiments based on the synthetic workloads and the real world traces validated the performance of FESTAL. Zhu et al. \cite{zhu2016fault} have utilized battery back-up scheduling schemes to establish an algorithm for fault-tolerant execution of scientific workflows (FASTER) by incorporating task allocation and message transmission features that employed a backward shifting approach for use of physical resources. A real-time workflow fault-tolerant model is developed which  extends the traditional primary backup model by including
 the cloud characteristics. Using this model, a task allocation and message transmission approach is developed to ensure fault tolerance during  information processing and  workflow execution. This algorithm enabled full utilization of idle resources and incorporated task overlapping and VM migration for improved resource utilization. Further, it applied the vertical/horizontal
 scaling-up technique to provision the resources for a bursty workflows, and used a vertical scaling-down scheme to avoid
 unnecessary and ineffective resource changes due to fluctuating workflow requests.
 \par The existing works have attempted to provide HA either by proactive methods such as resource failure prediction or reactive methods including  multi-cloud and running multiple replicas of a VM leading to high operational cost \cite{jhawar2012fault}. In contrast, SRE-HM furnishes both reactive as well as proactive approaches to maximize MTBF and minimize MTTR and thereby upgrading the service availability at both ends. Also, it reduces the major issue of providing replicas for all the VMs which leads to high operational costs.   SRE-HM estimates significance ranks of VMs according to the characteristics of tasks executed on them followed by a selection of the most optimal HA strategy corresponding to each VM's significance and user-specified constraints to enhance the HA experience for users with an optimized operational cost for CSPs. Moreover, the proposed model imparts a prior estimation of resource contention and mitigation of its worse effects by migrating VMs and  balancing load in real-time. The prediction of resource usage assists in alleviating server over-/under-load while reducing resource and power wastage. Table \ref{table:notation} shows the list of symbols with their explanatory terms used throughout the paper.
\begin{table}[htbp]
	\centering
	
	\caption[Table caption text] {Notations and their descriptions}  
	\label{table:notation}
			\begin{tabular}{|l|}
				\hline
							
				$U$: user; 				
				$M$:  number of users; 			
				$V$: virtual machine ; \\  \hline
				$Q$: number of VMs; 
 				$PM$: physical machine; 
 				$\mathds{AV}$: Availability Score;\\  \hline
 				 $P$: number of servers;   	
                $i$: index for servers; 
                $j$: index for VMs;\\ \hline 
                $k$: index for User; 
                $\omega_{kji}$: Mapping of $k^{th}$ user's $j^{th}$ VM on $i^{th}$ server;\\ \hline
				$\mathchorus{RU}$: resource utilization; 
				 $\mathchorus{PW}$: power consumption; $\mathchorus{N}$: VM as a node;
			  \\  \hline $r$:  user's request;
			  $\mathds{R}$: resources;
			   $\mathchorus{C}$: Cluster; $Mem$: Memory; 
				 $C$: CPU;\\  \hline 
				$\mathds{N}$: Number of resources;				 
				$\eta$: probability of server failure;\\ \hline
				$\mathchorus{E}_{ab}$: Edge between node a and b;
			$\mathchorus{fq}$: frequency of invocation;\\ \hline $\Omega$: mapping for HA selection;			
			$\mathchorus{S}(\mathchorus{N}_{a})$: Significance value of node `a';\\  \hline
			
		$\mathds{V}_{fp}$: Failure-prone VMs; 
		$\mathchorus{F}$: Failure probability of HA strategy;\\ \hline
				
				\hline
			\end{tabular}
\end{table}

%
%
%

\section{SRE-HM Model}
\begin{figure*}[!htbp]
	\centering
	\includegraphics[width=1.0\linewidth]{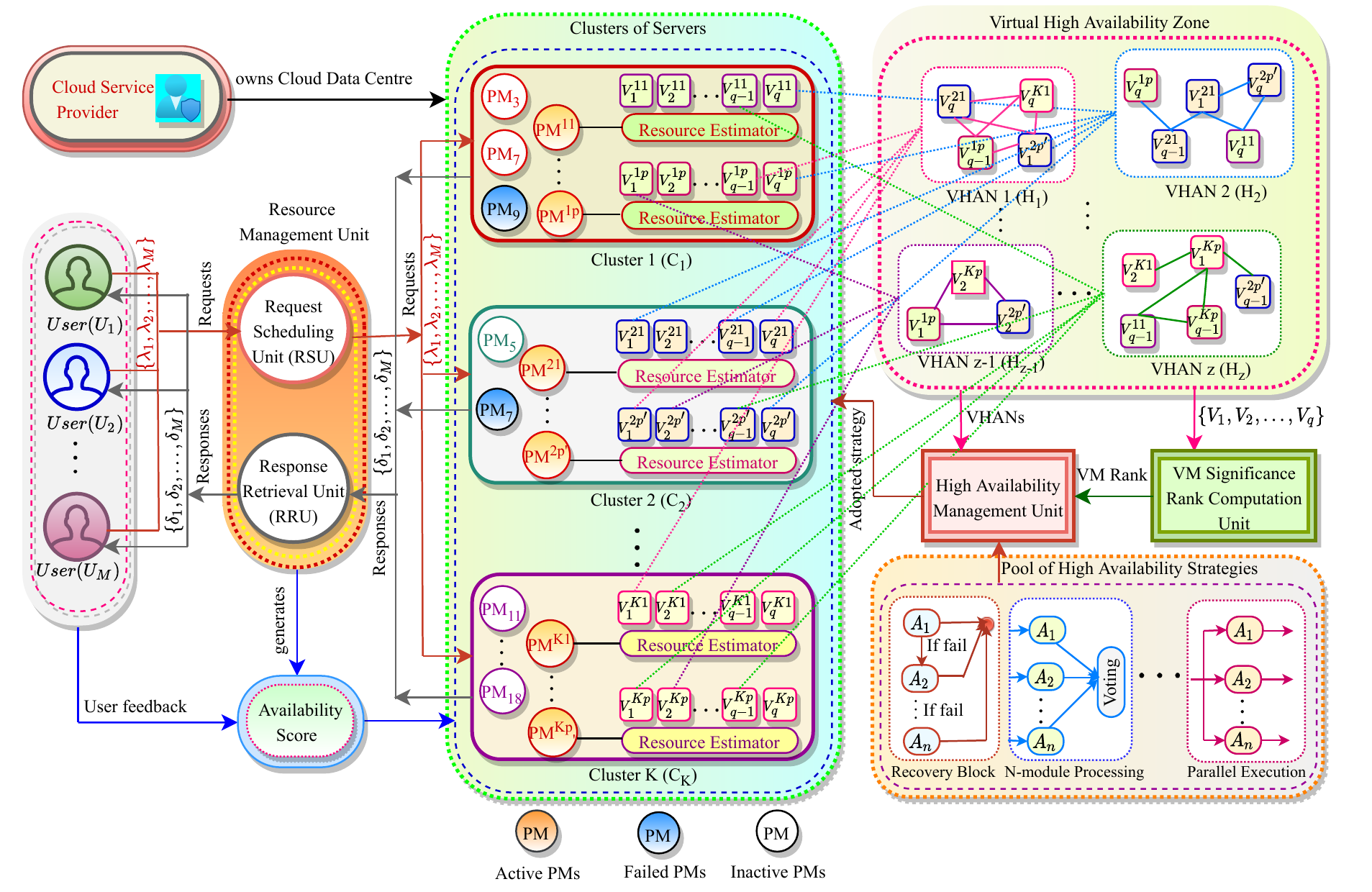}
	\caption{SRE-HM Model}
	\label{fig:proposedmodel}
\end{figure*}

Consider a Cloud Service Provider (CSP) owns a Cloud Data Centre (CDC) consisting of $K$ \textit{Clusters of Servers} (CoS) \{$C_1$, $C_2$, ..., $C_K$\} $\in \mathds{C}$ employing $P$ {physical machines (PM)} \{$PM^{k1}$, $PM^{k2}$, ..., $PM^{kp}$\} $\in \mathds{PM}$, where $ \forall k \in [1, K]$ as demostrated in Fig. \ref{fig:proposedmodel}. The PMs can be in any of the three possible states: \textit{active},  \textit{inactive}, or \textit{failed} due to hardware failure. Each active server deploys \textit{Resource Estimator} (RE) and $Q$ VMs  such that \{$V^{ki}_1$, $V^{ki}_2$, ..., $V^{ki}_q$\} $\in \mathds{V}$ are hosted on $i^{th}$ server $PM^{ki}$ within cluster $C_k$, where $\forall_i \in [1, P]$, $\forall_k \in [1, K]$.
 RE predicts resource contention or overloading condition on its associated active server ($PM$) to mitigate its worse effects proactively (discussed in Section \ref{RE}). CDC employs \textit{Resource Management Unit} (RMU) composed of two major sub-units \textit{Request Scheduling Unit} (RSU) and \textit{Response Retrieval Unit} (RRU) for distribution of users assigned job requests, retrieval of responses, and efficient management of physical resources.  Assume that $M$ \textit{Users} \{$U_1$, $U_2$, ..., $U_M$\} $\in \mathds{U}$ submit \textit{job requests} \{$\lambda_1$, $\lambda_2$, ..., $\lambda_M$\} to RSU for execution. Each request is a high computing application in the form of Bag of Tasks (BoT) such as \{$t^m_1$, $t^m_2$, ..., $t^m_b$\} $\in \lambda_m: \forall m \in [1, M]$, where $b$ is total number of tasks with a specific cost ($\mathchorus{EC}^m$) and deadline of execution ($\mathchorus{T}^m$). 

\par RSU distributes tasks \{$t^m_1$, $t^m_2$, ..., $t^m_b$\} $\in \lambda_m: \forall m \in [1, M]$ for execution among the VMs deployed on different PMs with in CoS. Thereafter, the responses \{$\delta_1$, $\delta_2$, ..., $\delta_M$\} generated by VMs are assembled and fetched to users \{$U_1$, $U_2$, ..., $U_M$\} with the help of RRU. Both RMU and users \{$U_1$, $U_2$, ..., $U_M$\} send consecutive feedback to CoS and \textit{High Availability Management Unit} (HAMU) by  evaluating the \textit{Availability Score} ($\mathds{AV}_{score}$) using the following Eq. (\ref{eq: availability score}); where $\mathds{AV}_g$ is guaranteed availability (as defined in SLA terms and conditions) and $\mathds{AV}_o$ is offered availability (actual availability of resources experienced by the user).        
\begin{equation}
\mathds{AV}_{score}=\frac{\mathds{AV}_g -\mathds{AV}_o}{\mathds{AV}_g}	\times 100 \label{eq: availability score}
\end{equation}

The availability score ($\mathds{AV}_{score}$) is sent periodically as a feedback for analysing the performance of currently adopted availability strategies and further improve the status of high availability management within CoS. The VMs executing tasks \{$t^m_1$, $t^m_2$, ..., $t^m_b$\} $\in \lambda_m: \forall m \in [1, M]$ of $m^{th}$ user have correlation among them because of mutual data exchange for successful execution of $m^{th}$ job request  $\lambda_m$. Therefore, a virtual network of such collaborating VMs is created depending on the scheduling of tasks (belonging to a common job request) on different VMs which are allocated under the constraints of high availability forming a \textit{Virtual High Availability Network} (VHAN). The proposed model introduces VHAN, Virtual High Availability Zone (VHAZ) and VM Significance Rank (VSR) which are defined as follows:

\begin{definition}[VHAN]
	A virtual network of VMs $\{V^m_1$, $V^m_2$, ..., $V^m_b\}$ having heterogeneous resource $\{\mathchorus{C}$, $\mathchorus{Mem}\}$ capacities executing inter-dependent tasks $\{t^m_1$, $t^m_2$, ..., $t^m_b\} \in \lambda_m: \forall m \in [1, M]$ deployed on different physical machines by the CSP, are  dedicated to provide high availability service to user $U_m$.
	
\end{definition}      
 
\begin{definition}[VHAZ]
	A virtual zone is a collection of multiple VHANs confined to provide high availability to the number of users $\{U_1$, $U_2$, ..., $U_M\}$ by employing selected optimal high availability strategies. 
\end{definition} 
\begin{definition}[VSR] A significance rank assigned to a VM is an estimated value for a quantitive measurement of its intendment based on the criticality and characteristics of the task it holds, frequency of invocations, and number of invocations by other VMs.   
\end{definition}

A VHAZ comprising of VHANs: $H_1$, $H_2$, ..., $H_M$ dedicated for users $U_1$, $U_2$, ..., $U_M$, respectively is designed, where a set of VMs \{$V^{21}_q$, $V^{K1}_q$, $V^{1p}_{q-1}$, $V^{2p'}_1$\} $\in H_1$;  \{$V^{1p}_q$, $V^{21}_1$, $V^{21}_{q-1}$, $V^{11}_{q}$, $V^{2p'}_q$\} $\in H_2$ and so on. The different VMs \{$V^{ki}_1$, $V^{ki}_2$, ..., $V^{ki}_{b}$\} $\in H_m$: $\forall k \in [1, K], i \in [1, q], m \in [1, M]$ constituting $m^{th}$ VHAN are physically deployed on different PMs within CoS. The essential constraints that must be satisfied for VM deployment on a PM are stated in Eq. (\ref{VMP}), where $\omega_{kji}$ represents a mapping $\omega_{kji}: C_k \times V_j \times PM_i \in \{1, 0\}$ such that $\omega_{kji}= 1$ if VM $V_j$ is deployed on server $PM_i$ within cluster $C_k$, else, it is $0$;  $\forall i \in [1, P], k \in [1, K]$;  ${\mathds{R}}$ specifies resources viz., CPU ($\mathchorus{C}$) and memory ($\mathchorus{Mem}$) for assignment of VM ($V_j$) on $PM_i$. 
\begin{gather}
\sum_{j=1}^{Q}{V^{ki}_j\times {{\mathds{R}_j}} \times \omega_{kji} \le PM_i\times {{\mathds{R}_i}}}  \label{VMP}
\end{gather}   
The resource utilization of all the VMs belonging to each VHAN is periodically monitored and the probability of failure is estimated proactively with the help of RE. The significance ranks of VMs (i.e., VSR) belonging to different VHAN are estimated by employing  \textit{VM Significance Rank Computation Unit} (VSRCU) which is discussed in detail in Section \ref{vsr}. VSR helps in determining the valuable contributions of different VMs for the successful execution of a job request $\lambda$ in a VHAN and thereby allows an optimal selection of a high availability strategy from the pool encapsulating different HA strategies. HAMU fetches information from VSRCU, VHAZ, and a pool of HA strategies (Section \ref{ha selection}) to decide and adopt the most suitable HA strategy for each VM in VHAN to implement the best possible high availability environment for requests execution. 

 \section{PM Failure Estimation and Handling} \label{RE}
\begin{figure*}[!htbp]
 	\centering
 	\includegraphics[width=0.75\linewidth]{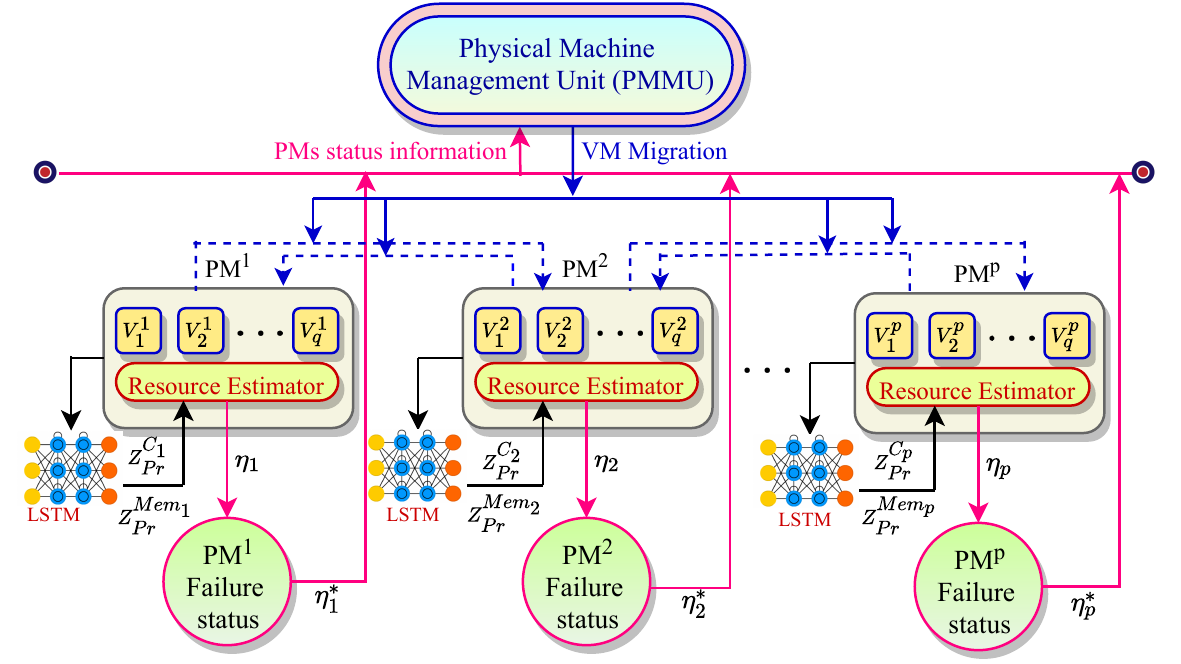}
 	\caption{Mechanism for Physical Machine Availability }
 	\label{fig:pmestimator}
 \end{figure*}

 The availability of physical machines is managed with the help of a Resource Estimator (RE) dedicated to analyse the upcoming resource demand on a particular server proactively. Fig. \ref{fig:pmestimator} portrays a complete mechanism for managing the availability of PMs, where RE utilizes Long Short Term Memory (LSTM) Neural Network for accurate prediction of expected resource usage on a server in the future which is trained/re-trained periodically.  LSTM neural network is chosen for server failure prediction due to their high capability of remembering information for a longer duration and extracting intuitive patterns by finding useful correlations among them \cite{kumar2018long}.
  The physical machines $PM^1$, $PM^2$, ..., $PM^p$ employ an exclusive LSTM neural network based resource predictor, optimized with the latest historical resource usage information of VMs. 
LSTM is a cell state, where  previous block ($\mathchorus{S}^{t-1}$) information flows  to the current block ($\mathchorus{S}^t$). It is composed of four neural network layers, where the first layer ($\mathchorus{G}_1$) decides the amount of previous information of resource usage ($\mathchorus{cf}_{\mathchorus{RU}_i}^t$) to be passed using Eq. (\ref{lstm1}), where $\mathchorus{W}$ is the weight matrix, $\mathchorus{b}$ is a bias value, $Z_{Pr}^{\mathchorus{RU}^{t-1}}$ and $\mathchorus{RU}_i^t$ are previous output and current input, respectively. The cell state is updated using two network layers viz., $sigmoid$ layer ($\mathchorus{G}_2$) and $tanh$ layer. $\mathchorus{G}_2$
decides the values to be updated ($\mathchorus{I}^t$) using Eq. (\ref{lstm2}) and $tanh$ layer generates a new candidate values vector ($\hat{\mathchorus{S}}^t$) as stated in Eq. (\ref{lstm3}). Finally, Eq. (\ref{lstm4}) combines both outputs to update cell state.  
\begin{gather}\label{lstm1}
\mathchorus{cf}_{\mathchorus{RU}_i}^t = \mathchorus{G}_1(\mathchorus{W}_{\mathchorus{cf}}\cdotp [Z_{Pr}^{\mathchorus{RU}^{t-1}}, \mathchorus{RU}^t] +\mathchorus{b}_{\mathchorus{RU}})
\\ \label{lstm2}
\mathchorus{I}^t = \mathchorus{G}_2(\mathchorus{W}_{\mathchorus{I}}\cdot [Z_{Pr}^{\mathchorus{RU}^{t-1}}, \mathchorus{RU}^t] +\mathchorus{b}_{\mathchorus{I}})
\\ \label{lstm3}
\hat{\mathchorus{S}}^t = \mathchorus{tanh}(\mathchorus{W}_{\mathchorus{S}}\cdot [Z_{Pr}^{\mathchorus{RU}^{t-1}}, \mathchorus{RU}^t] +\mathchorus{b}_{\mathchorus{S}})
\\ \label{lstm4}
\mathchorus{S}^t = \mathchorus{cf}_{\mathchorus{RU}_i}^t \times \mathchorus{S}^t  + \hat{\mathchorus{S}}^{t-1} \times \mathchorus{I}^t 
\end{gather}
This LSTM based resource predictor periodically forecasts CPU ($Z_{Pr}^{\mathchorus{C}_i}$) and memory usage ($Z_{Pr}^{\mathchorus{Mem}_i}$) for $PM^i$ to be provided to RE for estimation of the probability of failure ($\eta_i^\ast$) of the respective PM using Eq. (\ref{eq: RE}), where $\mathchorus{RU}_i^{Thr}$ is a threshold value of resource utilization viz., $\mathchorus{C}$ and $\mathchorus{Mem}$ usage value for $PM^i$ and $\forall_i \in [1, P]$. This threshold value is decided by RMU in the real world cloud environment and it is set to 85\% for the empirical evaluation. 
 \begin{equation} \label{eq: RE}
 	\eta_i^\ast=	\begin{cases}
 	1, & {If(\mathchorus{RU}_i^{Thr} \ge \eta_i)} \\
 
 	0, & {\text{otherwise}}  
 	\end{cases} \quad \eta_i: \cup(Z_{Pr}^{\mathchorus{C}_i}, Z_{Pr}^{\mathchorus{Mem}_i}) 
 \end{equation}

A  \textit{Physical Machine Management Unit} (PMMU) is employed for proactive handling of any resource-contention based failure  by allowing VM migration. PMMU gathers failure information of PMs as \{$\eta_1^\ast$, $\eta_2^\ast$, ..., $\eta_p^\ast$\} and alleviates them by migrating VMs before occurrence.  The largest size VMs is migrated (to avoid the occurrence of resource congestion frequently) from the estimated overloaded PM
 to a suitable active PM preferably rather than waking up an inactive PM for minimization of overall power consumption in CDC. The inactive PMs are turned ON only when currently activated PMs are unable to host the migrating VMs. 
 \par If the predicted information of resource usage of a VM is higher or lesser than the current resource usage of that VM, then the respective over-utilization or under-utilization of the PM is estimated. In case of probable over-utilization of the PM, the chances of resource congestion increase which indicates the probability of PM failure (i.e., $\eta^\ast$ becomes TRUE). Accordingly, the  status of VM (${V}_j^{{status}}$) turns into `1' in Eq. (\ref{mig}) which triggers the migration of the respective VM to alleviate the effect of server over/under-load before their actual occurrence. The  VM migration cost ($Mig_{cost}$) is computed using Eq. (\ref{mig_cost}).  

\begin{gather}\label{mig}
{{V}_j}^{status}=\begin{cases}
1  & {If(\omega_{kji} \times \eta_i^\ast = 1)} \\
0  & {\text{otherwise.}} 
\end{cases}
\\ \label{mig_cost}
\resizebox{0.48\textwidth}{!}{$  
	Mig_{cost}={(\sum{c_{mig.j}*(D(PM_i ,PM_{ii})\times W(V_{mig}))}) + \sum{n_{i}\times {d_{i}}}}
	$}
\end{gather}

where $D(PM_i ,PM_{ii})$ is the distance or number of hops covered by $V_{mig}$ from source ($PM_i$) to destination server $PM_{ii}$, \{$i, ii \in [1,P], i \neq ii$\}, $V_{mig}$, $W(V_{mig})$ = $V_{mig}^{\mathchorus{C}} \times V_{mig}^{\mathchorus{M}}$ is the size of migrating VM. The first term $\sum{c_{mig.{ii}}*{D(PM_i ,PM_{ii}) *W(V_{mig})}}$ specifies energy consumed during VM migration. The second term $\sum{n_{i}*{d_{i}}}$ states energy consumed in the server state transition, where if $j^{th}$ VM is placed at ${ii}^{th}$ server after migration then $c_{mig.{ii}}=1$, otherwise, $c_{mig.{ii}}=0$. If the ${ii}^{th}$ server receives one or more VMs after migration, then $n_{ii}=1$ else it is 0. Similarly, if $d_{ii}=0$ then ${ii}^{th}$ server is already active before migration, otherwise, $d_{ii}=E_{tr}$ where $E_{tr}$ is the energy consumed in switching a server from sleep to active state. 
\section{Significance Rank Computation} \label{vsr}

A group of inter-communicating VMs in a VHAN ($H_m$) executing a cloud application: \{$t^m_1$, $t^m_2$, ..., $t^m_n$\} $\in \lambda_m$: $\forall m \in [1, M]$, is considered as a weighted directed graph (WDG), where a VM represents a node ($\mathchorus{N}_a$) and invoking relation between VMs is a directed edge ($\mathchorus{E}_{ab}$) from $\mathchorus{N}_a$ to $\mathchorus{N}_b$ in WDG. A weight value ($\xi(\mathchorus{E}_{ab})$) is assigned to each edge in WDG using Eq. (\ref{eq:ew}), where $\mathchorus{fq}_{ab}$ is the invocation frequency of node $\mathchorus{N}_a$ by node $\mathchorus{N}_b$, $n$ is the number of nodes in WDG, $\mathchorus{C}_a $ determines whether $\mathchorus{N}_a$ is critical ($\mathchorus{C}$) or not, which is $1$ for $\mathchorus{N}_a$ is a critical node, else it is $0$, and if $\mathchorus{N}_a$ invokes $\mathchorus{N}_b$ then  $\mathchorus{fq}_{ab}= 1$; otherwise, $\mathchorus{fq}_{ab}= 0$. 
\begin{equation}
\label{eq:ew}
\xi(\mathchorus{E}_{ab})=\frac{\mathchorus{fq}_{ab} \times \mathchorus{C}_a}{\sum_{b=1}^{n}{\mathchorus{fq}_{ab}}}
\end{equation}  
Hence, the edge $\mathchorus{E}_{ab}$ has a higher weight value if $\mathchorus{N}_a$ is invoked more frequently by $\mathchorus{N}_b$ as compared to other nodes invoked by $\mathchorus{N}_b$. WDG containing $n$ nodes is represented as $n \times n$ matrix $\xi$ established using Eq. (\ref{eq:ew}) subject to Eq. (\ref{eq:ew1}), where $\xi(\mathchorus{E}_{ab})=\frac{1}{n}$ if $\mathchorus{N}_a$ has all the incoming edges only. 
\begin{equation} \label{eq:ew1}
\forall a, \sum_{b=1}^{n}{\xi(\mathchorus{E}_{ab})}=1
\end{equation}
Initially, a random value in the range [0, 1] is assigned to all the nodes of WDG. All the nodes are differentiated on the basis of their characteristics into a critical node ($\mathchorus{C}$) and a non-critical node ($\mathchorus{NC}$). Accordingly, the significance value of node $\mathchorus{N}_a$ represented as $\mathchorus{S}(\mathchorus{N}_a)$ is estimated using either Eq. (\ref{eq:vsr1}) or Eq. (\ref{eq:vsr2}) for critical or non-critical nodes, respectively;
\begin{gather} \label{eq:vsr1}
\mathchorus{S}(\mathchorus{N}_a)=
(1-\mathchorus{d}){\frac{\Psi}{|\mathchorus{C}|}+ \mathchorus{d}{\sum_{g\in Z(\mathchorus{N}_a)}{\mathchorus{S}(\mathchorus{N}_g)\xi(\mathchorus{E}_{ga})} }}
\\ \label{eq:vsr2}
\mathchorus{S}(\mathchorus{N}_a)=
(1-\mathchorus{d}){\frac{1-\Psi}{|\mathchorus{NC}|}+ \mathchorus{d}{\sum_{g\in Z(\mathchorus{N}_a)}{\mathchorus{S}(\mathchorus{N}_g)\xi(\mathchorus{E}_{ga})} }}
\end{gather}
where $|\mathchorus{C}|$ and $|\mathchorus{NC}|$ are number of critical and non-critical nodes, respectively such that $|\mathchorus{C}|+|\mathchorus{NC}|=n$, $Z(\mathchorus{N}_a)$ is a set of nodes invoking $\mathchorus{N}_a$, $d$ is an adjusting parameter in the range [0, 1]. The parameter $\Psi$ ($\frac{|\mathchorus{C}|}{n} \le \Psi \le 1$) determines the effectiveness of critical and non-critical components in the estimation of VSR as when $\frac{|\mathchorus{C}|}{n} < \Psi \le 1$, the value of critical node $(1-\mathchorus{d}){\frac{\Psi}{|\mathchorus{C}|}}$ is greater than non-critical node $(1-\mathchorus{d}){\frac{1-\Psi}{|\mathchorus{NC}|}}$. If $\Psi$=1, then value of the critical nodes is 1, and larger significant values are assigned to critical nodes, while if $\Psi=\frac{|\mathchorus{C}|}{n}$, both the critical and non-critical nodes show equal effects. Finally, the VMs are ranked in the descending order of their significance values.  VSRCU estimates the significance value based ranks of VMs using an enhanced version of the Weighted Page Rank algorithm \cite{xing2004weighted} as mentioned above.  

\section{HA Strategy Selection}\label{ha selection}
The user specifies several high availability constraints regarding response time and total cost of application execution in the SLA to be requited during actual operation. HAMU incorporates significance ranks of VMs ($\mathchorus{S}$), response time ($\mathchorus{T}$), failure probability of $j^{th}$ HA strategy ($\mathchorus{F}_j$), and execution cost ($\mathchorus{EC}$) which yields a mapping $\Omega$ defined in Eq. (\ref{eq:ha}) to select most admissible HA strategy ($Y_j$) for VM ($V^m_i$) of $m^{th}$ user subject to constraints stated in Eqs. (\ref{c1}), (\ref{c2}), (\ref{c3}), where $D$ is total number of HA strategies.
\begin{gather}\label{eq:ha}
\Omega: \sum_{j}^{D}\mathchorus{S}(V^m_i) \times Y_j \times \mathchorus{F}_j \quad \forall_i \in [1, n] \quad s.t.  \\
\sum_{j}^{D} Y_j \times \mathchorus{EC}_j \le V^m_i \times \mathchorus{EC}^m_i \times U_m, \label{c1}  \\
\sum_{j}^{D} Y_j \times \mathchorus{T}_j \le V^m_i \times \mathchorus{T}^m_i \times U_m, \label{c2}\\
\sum_{j}^{D} Y_j =1, \quad Y_j \in \{0, 1\} \label{c3} 	
\end{gather}
In this work, \textit{Automatic recovery block} (ARP), \textit{Multi-version programming} (MVP), and \textit{Parallel execution} (PE) based HA strategies are adopted:
\begin{itemize}
	\item \textit{ARP}: It engages redundant standby images of a VM which are invoked in sequence if currently an active image of VM fails. ARP failure probability ($\mathchorus{F}^{ARP}$) is computed using Eq. (\ref{eq:arp}), which get fail only if all the replicated images get fail, where $num$ is the number of replicated images and $\mathchorus{F}^{ARP}_{i}$ is failure  of $i^{th}$ VM.  
	\begin{equation} \label{eq:arp}
		\mathchorus{F}^{ARP}= \prod_{i=1}^{num} \mathchorus{F}^{ARP}_{i}
	\end{equation}
	\item \textit{MVP}: It appoints multiple active images or versions of a VM instance concurrently for the execution of a common task and the final output is obtained by majority voting. Eq. (\ref{eq:mvp}) computes its failure probability, where $num$ is a number of versions ($num$ is an odd number), and $\mathchorus{f}(i)$ is the failure probability of alternative VM images. MVP fails only if more than half of redundant VM images get fail.
	   \begin{equation} \label{eq:mvp}
	   \mathchorus{F}^{MVP}= \sum_{i=\frac{num+1}{2}}^{num} \mathchorus{f}(i)
	   \end{equation}
	  \item \textit{PE}: It allows parallel execution of all $num$ versions of $i^{th}$ VM instance and the first generated response is provided as final output. It fails only if all the parallel executions get fail and its failure probability is computed in Eq. (\ref{eq:pe}) 
	  \begin{equation} \label{eq:pe}
	  \mathchorus{F}^{PE}= \prod_{i=1}^{num} \mathchorus{F}^{PE}_{i}
	  \end{equation}
\end{itemize}

 \section{Operational Design and Complexity}
The proposed model initializes a list of PMs, VMs, and users, where the users submit requests for  execution of applications to the resource manager module which assigns these requests on different VMs placed on PMs as stated in Algorithm \ref{algo-RP-SVS}. For each time-interval \{$t1$, $t2$\}, VMs resource utilization is predicted using LSTM, which further helps to determine any overload or resource-contention problem and mitigate this situation by VM migration, proactively. The user becomes an ephemeral
owner of a group of VMs engaged in task execution belonging to the respective user's application, thus creating a VHAN. On the same line, for each user exclusive VHAN is generated, which are later transformed into a WDG matrix. An optimal HA strategy is selected by including only those strategies which satisfies user specified constraints, among which a HA strategy with the least failure probability is selected as the optimal strategy for the target VM. 
 \begin{figure}[!htbp]
 	\removelatexerror
 	\begin{algorithm}[H]
 		\caption{SRE-HM Operational Summary}
 		\label{algo-RP-SVS}
 		Initialize: $List_{{\mathds{S}}}$, $List_{\mathds{V}}$, $List_{\mathds{U}}$\; 
 		
 		\For {each time-interval $\{t_1, t_2\}$}{ 
 			
 			{Predict $\mathchorus{RU}$ on each VM on a PM and aggregate it to analyse resource-contention  \;
 				Migrate maximum size VM from overloaded PM by applying the steps for PMMU in Section \ref{RE} \;
 				
 			}
 			User submitted jobs are distributed into tasks to be executed on VMs (arranged in VHAN) placed on different PMs while satisfying resource constraints (Eq. (\ref{VMP}))\;
 			Estimate VSR for each VM in VHAN by applying concepts mentioned in Section \ref{vsr}\;
 			WDG in the form of $n\times n$ matrix is built from VHAN \;
 			\For {($j=1$; $j \le D$; $j++$)}{
 			\If{Eq. (\ref{c1}) \&\& Eq. (\ref{c2}) \&\& Eq. (\ref{c3}) }{$x_j= \mathchorus{F}_j$\;}
 			{continue\;}
 		
 	}
 		Select HA strategy which has minimal failure probability among all $x_j$\;	 
 		}

 	\end{algorithm}
 	
 \end{figure}

 Step 1 initializes lists of servers, users, and VMs associated with different users and consumes time-complexity $O(1)$. Let steps 2-15 repeat for $t$ time-slots, where step 3 appoints LSTM-RNN for prediction of resource utilization with complexity of $O(\mathchorus{h})$, where $\mathchorus{h}$ is length of input sequence. Step 4 performs VM migration has $O(\mathchorus{m})$ complexity, where $\mathchorus{m}$ is a number of migrations. Steps 5-7 create VHAN followed by WDG in the form of $n\times n$ matrix have $O(n^2)$ computational complexity. Steps 8-13 iterate $D$ times with $O(D)$ complexity while Step 14 shows $O(1)$ complexity. Hence, overall computational complexity turn into $O(\mathchorus{h}\mathchorus{m}n^2Dt)$.

 \section{Performance Evaluation}
  
 \subsection{SRE-HM Implementation}
 A SRE-HM prototype is configured with the collaboration of major modules discussed below:
 \begin{itemize}
 	\item \textit{VMs Resource Estimation}: The future resource usage of different VMs hosted on each server is predicted and their number and size are scaled accordingly. Also, any overload or resource contention failure is detected based on prediction and mitigated by VM migration proactively.
 	\item \textit{VMs Allocation}: The predicted VMs are assigned to the servers conforming to the resource distribution constraints specified in Eq. (\ref{VMP}).
 	\item \textit{User's Task submission}: Each user submits different tasks along with their deadline and cost of execution, to be assigned to VMs hosted on different servers. 
 	\item \textit{Generation of WDG}: A WDG is built for each user by establishing links and invocations randomly among different VMs   selected for task execution of the  user.
 	\item \textit{Deciding Critical and Non-Critical VMs}: Based on the number of invocations of a VM, greater than a threshold value of invocation (which is set to 3 in the experiments), are taken as `critical VMs' and remaining VMs are `non-critical VMs'.
 	\item \textit{VM Ranks Computation}: The VM ranks are estimated  for critical and non-critical VMs exclusively by following the steps mentioned in Section (\ref{vsr}).
 	\item \textit{HA Strategy Selection and execution}: As per the computed ranks of VMs, failure probability of different HA strategies, cost and deadline of task execution, an admissible strategy with the least perhaps failure probability is selected for each VM. This module is extensible, where any ranking algorithm and HA strategies can be incorporated adaptively.
 	\item \textit{Availability and Execution Cost Evaluation}: Based on the above modules, the availability is computed using Eq. (\ref{availability}) while the cost is estimated in terms of resource utilization and power consumption computed as given below.  
 		
 \end{itemize}
\subsection{Parameters evaluation}
\begin{figure}[!htbp]
	\centering
	\includegraphics[width=0.99\linewidth]{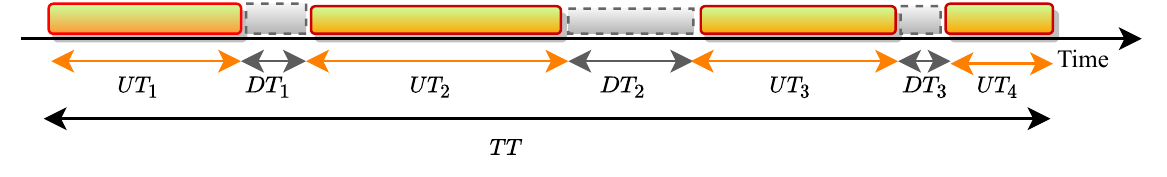}
	\caption{Lifecycle of a service indicating  uptime, outage and total time}
	\label{fig:mttrmtbf}
\end{figure}

The HA performance metrics including MTBF, MTTR, and average availability of SRE-HM model are evaluated by 
considering lifecycle of a hypothetical service  as demonstrated in Fig. \ref{fig:mttrmtbf}, wherein the variables $DT$, $UT$, and $TT$ are downtime, uptime, and total time of a service, respectively. 
A service may be either in uptime as illustrated by the variables $UT_1$, $UT_2$, $UT_3$, and $UT_4$ or in downtime defined by variables $DT_1$, $DT_2$, and $DT_3$. The term $Num_{\mathchorus{F}}$ represents a number of failures of system (i.e., 3 in given Fig. \ref{fig:mttrmtbf}); the MTBF and MTTR are evaluated using Eqs. (\ref{mtbf}) and (\ref{mttr}), respectively. MTTF and MTTR are stated as the averages of uptime and downtime, respectively in these equations. Accordingly, the average availability can be computed using  Eq. (\ref{availability}), where $Num_{\mathchorus{F}}$ is total number of failures, $\sum_{i=1}^{M}{UT_i}$ and $\sum_{i=1}^{M}{DT_i}$ represent total uptime and downtime of a service, respectively experienced by $M$ users over time-interval \{$t_1$, $t_2$\}.

\begin{equation}
\label{mtbf}
MTBF=\int\limits_{\substack{t_1\\\mathcal{}}}^{t_2}(\frac{\sum_{i=1}^{M}{UT_i}}{Num_{\mathchorus{F}}})dt
\end{equation}
\begin{equation}
\label{mttr}
MTTR= \int\limits_{\substack{t_1\\\mathcal{}}}^{t_2}(\frac{\sum_{i=1}^{M}{DT_i}}{Num_{\mathchorus{F}}})dt
\end{equation}
\begin{equation}
\label{availability}
A_{avg} = \frac{MTBF}{MTBF+MTTR} 
\end{equation} 
 The resource utilization ($\mathchorus{RU}$) of data centre is estimated using Eqs. (\ref{RU1}) and (\ref{RU2}), where $\mathds{N}$ is the number of  resources, $\mathchorus{RU}^\mathchorus{C}$ and $\mathchorus{RU}^\mathchorus{Mem}$ are CPU and memory of server. If $k^{th}$ server $PM_k$ is active, hosting VMs ($\beta_k$ = 1), otherwise, $\beta_k$ = 0.  
 \begin{equation}
 \mathchorus{RU}^{DC}= \frac{\sum_{k=1}^{P}{\mathchorus{RU}_k^{\mathchorus{C}}} + \sum_{k=1}^{P}{\mathchorus{RU}_k^\mathchorus{Mem}}}{|\mathds{N}| \times \sum_{k=1}^{P}{\beta_k} } \label{RU1}
 \end{equation} 
 \begin{equation}
 \mathchorus{RU}_k^{\mathds{R}} = \frac{\sum_{i=1}^{Q}{\omega_{ik}} \times V_i^{\mathds{R}}}{S_k^{\mathds{R}}} \quad \forall_k \in \{1, P\}, \mathds{R} \in \{\mathchorus{C}, \mathchorus{Mem}\} \label{RU2}	
 \end{equation}
 Eq. (\ref{power2}) computes power consumption ($\mathchorus{PW}^{DC}$), where ${PW_i}^{max}$, ${PW_i}^{min}$, and ${PW_i}^{idle}$ are maximum, minimum, and idle state power consumption, respectively of $i^{th}$ server.
 \begin{equation}
 \mathchorus{PW}^{DC} = 
 \sum_{i=1}^{P} {[{PW_i}^{max} - {PW_i}^{min}]\times{RU} + {PW_i}^{idle}}
 \label{power2}
 \end{equation} 
 
 \subsection{Experimental Set-up and {Dataset}}
 The simulation experiments are executed on a server machine assembled with two Intel\textsuperscript{\textregistered} Xeon\textsuperscript{\textregistered} Silver 4114 CPU with 40 core processor and 2.20 GHz clock speed. The server machine is deployed with 64-bit Ubuntu 16.04 LTS, having main memory of 128 GB. The data centre environment included three different types of servers and four types of VMs configuration shown in Tables \ref{table:server} and \ref{table:vm} in Python. The resource features like power consumption ($PW_{max}, PW_{min}$), MIPS, RAM and memory are taken from real server IBM \cite{IBM1999} configuration where $S_1$ is `ProLiantM110G5XEON3075', $S_2$ is `IBMX3250Xeonx3480' and $S_3$ is `IBM3550Xeonx5675'. The VMs configuration is inspired from the VM instances of Amazon website \cite{amazon1999EC2}. 
 
 \begin{table}[!htbp]
 	\centering
 	
 	\caption[Table caption text] {Server Configuration}  
 	\label{table:server}
 	\resizebox{8.5cm}{!}{
 		\begin{tabular}{lccccc}
 			\hline
 			Server&PE&MIPS&RAM(GB)&$PW_{max}$&$PW_{min}$/$PW_{idle}$\\
 			\hline
 			$S_1$ 	& 2&2660&4&135&93.7 \\
 			$S_2$	& 4&3067&8&113&42.3 \\
 			$S_3$	& 12&3067&16&222&58.4 \\

 			\hline
 	\end{tabular}}
 \end{table}
 
 \begin{table}[!htbp]
 	\centering
 	
 	\caption[Table caption text] {VM Configuration}  
 	\label{table:vm}
 	\begin{tabular}{lccc}
 		\hline
 		VM type& PE &MIPS&RAM(GB)\\
 		\hline
 		$v_{small}$&1&500&0.5\\
 		$v_{medium}$&2&1000&1\\
 		$v_{large}$&3&1500&2\\
 		$v_{Xlarge}$&4&2000&3\\

 		\hline
 	\end{tabular}
 \end{table}
 \textit{Dataset}: Google Cluster Data (GCD) dataset is utilized for performance estimation of SRE-HM and comparative approaches which contains resources viz., CPU, memory, disk I/O  request and resource usage information of 672,300 jobs executed on 12,500 servers for the period of 29 days \cite{reiss2011google}. The CPU and memory utilization  percentage of VMs are obtained from the given CPU and memory usage percentage for each task in every five minutes over period of twenty-four hours.

 \subsection{Results}
 
 Table \ref{table:performanceGCD} reports the performance metrics: MTTR, MTBF, average availability ($\mathds{AV}$), accuracy of failure prediction ({$Acu^{Pr}$}), average number of overloads ($\mathds{OV}$), power consumption (\textit{$\mathchorus{PW}$}), Resource utilization ($\mathchorus{RU}$), average number of VM migrations ($\mathchorus{Mig}$) achieved for varying percentage of failure-prone VMs ($\mathds{V}_{fp}$) over period of 500 minutes. 
 \begin{table*}[!htbp]
 	
 	\caption[Table caption text] {Performance metrics for GCD workloads}  
 	\label{table:performanceGCD}
 	\small
 		\centering
 		\begin{tabular}{|l|c|c|c|c|c|c|c|c|c|}
 			\hline
 			
 		$\mathds{V}_{fp}$&$T(min.)$& $MTTR$ &$MTBF$&$\mathds{AV}$ &$Acu^{Pr}$ &$\mathds{OV}$ &$\mathchorus{PW}$ (W) &$\mathchorus{RU}$ &$ \mathchorus{Mig}\texttt{\#}$  \\ \hline \hline			
 			\multirow{3}{*}{5}
 	     	&100& 1.47& 2757.14 &99.91&99.3&5&8660.9&76.2 &7 \\ \cline{2-10}
 			&250&1.05 & 3900 & 99.96& 98.8& 7&8660.9&76.6 &5\\ \cline{2-10}
 			&300&1.05&3900&99.96&98.6& 7&8660.9&76.6 &5 \\ \cline{2-10}
 			&500& 0.84& 4900&99.98&99.6& 8&8660.9&76.5 &4\\ \hline \hline
 			
 			\multirow{5}{*}{10}
 			&100&2.73&1438.46& 99.84&97.9&11&8498.6&75.8 &13 \\ \cline{2-10}
 			&250&2.1&1900&99.85&98.4&9&8498.6&75.3 &10 \\ \cline{2-10}
 			&300&1.89&2122.22&99.81&95.5&11&8498.6&75.8  &9 \\ \cline{2-10}
 			&500&1.68 & 2400&99.93&98.7&8&8498.6&75.6 &8 \\ \hline \hline
 			
 			\multirow{5}{*}{30}
 			&100&3.99& 952.63&99.58&98.8&14&8601.7&77.8&19 \\ \cline{2-10}
 			&200&3.57& 1076.47&99.66& 97.6&16&8601.7&76.9 &17 \\ \cline{2-10}
 			&300&3.15& 1233.33&99.74  &98.8& 12&8601.7&76.7&15 \\ \cline{2-10}
 			&500& 2.94& 1328.57&99.78&93.7&19&8601.7&76.8&14 \\ \hline \hline
 			\multirow{5}{*}{80}
 			&100&4.62&809.09&99.43&97.9&15&8578.4&74.8 &22 \\ \cline{2-10}
 			&250&4.41& 852.38&99.48&98.2&18&8578.4&75.1 &21\\ \cline{2-10}
 			&300&3.99& 952.63&99.58&98.2&18&8578.4&75.1 &19 \\ \cline{2-10}
 			&500&3.36& 1150.00&99.76&98.8& 18&8578.4 &74.9&16 \\ \hline 
 			
 	\end{tabular}
 \end{table*}
The values achieved for MTBF and MTTR depends on the number of failures ($Num_{\mathchorus{F}}$) as depicted in Eqs. (\ref{mtbf}) and (\ref{mttr}), respectively. The values of uptime ($UT$) are obtained by calculating the product of {number of successfully deployed VMs}. The value of MTTR associated with a VM is 0.21 minutes which is reported in \cite{araujo2014availability}, \cite{santos2017analyzing}. Accordingly, the values of MTTR are enumerated for different number of VM migrations that changes with the number of unpredicted VM failures. The resultant values of availability  are evaluated using Eq. (\ref{availability}) depending on the MTBF and MTTR values recorded over time-interval \{$t_1$, $t_2$\} i.e., 100 minutes for the observed experiments. The availability for the Google Cluster workload is above 99\% for all the observed cases. 
 It can be observed that with increasing number of $\mathds{V}_{fp}$, the performance of SRE-HM (independent of time) is durable where  $Acu^{Pr}$, and $\mathchorus{RU}$ are greater than  97\%, and 74.5\%, respectively while  $\mathchorus{PW}$, and $\mathchorus{Mig}$ lesser than 8661 W and 22 VM migrations, respectively.
 
 \subsection{Comparison}
  
The different  versions  of SRE-HM including SRE-HM ($S^+$), SRE-HM with only critical VM ranking  ($S^\ast$), Without SRE-HM ($S^-$), and SRE-HM without VM ranking ($S^{\ast\ast}$) collaborated with three HA strategies ($arp$, $mvp$, $pe$) are implemented and compared for intermediate result values and performance metrics.  
\subsubsection{PM Failure Estimation}
\begin{figure}[!htbp]
	\centering	
	\subfigure[Failure Prediction Accuracy]{\includegraphics[width=0.69\linewidth, scale=2]{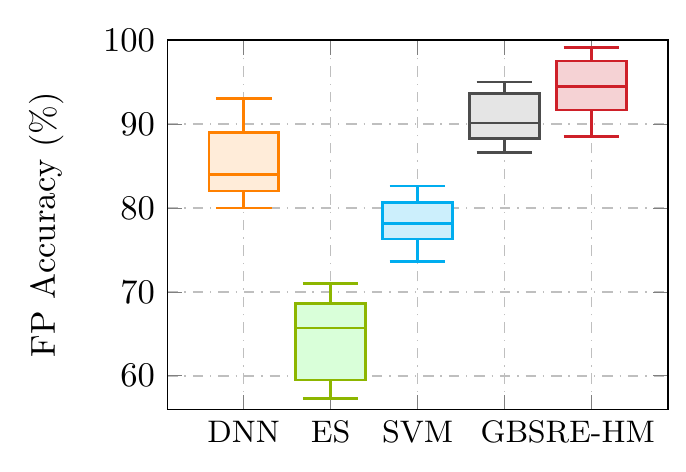}} 
	\subfigure[ PM Failures]{\includegraphics[width=0.69\linewidth, scale=2]{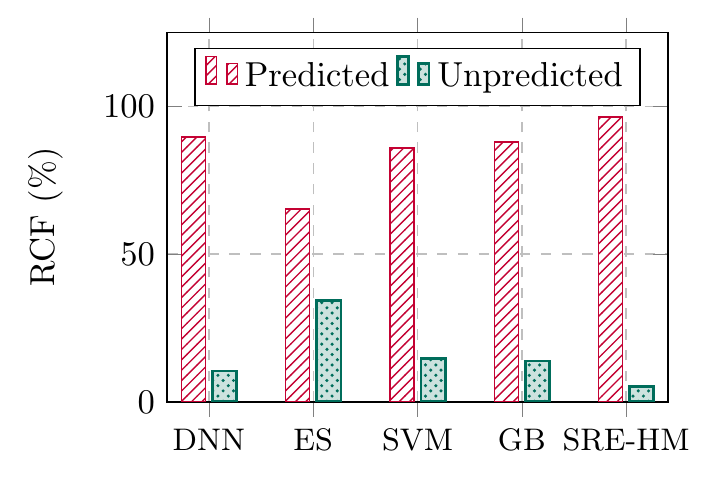}}
\caption{PM Failure Estimation}
	\label{ResourceEstimation}	
\end{figure}

Fig. \ref{ResourceEstimation}(a) compares boxplots for the outcomes of PM failure estimation
of SRE-HM with state-of-the-arts: DNN \cite{marahatta2020pefs}, ES \cite{sharma2019failure}, SVM \cite{pinto2016hadoop}, GB \cite{xu2018improving}. It shows that the upper, median, as well as lower quartiles have  highest value of prediction accuracy for SRE-HM prediction approach over all the compared methods. On the same lines, the percentage of resource contention based PM failures (RCF) is least for SRE-HM up to 72.2\% against ES as depicted in Fig. \ref{ResourceEstimation}(b).

\subsubsection{VM Ranking and HA Strategy Selection}
\begin{figure}[!htbp] 

	\centering	
	\subfigure[VM Ranking]{\includegraphics[width=0.69\linewidth, scale=2]{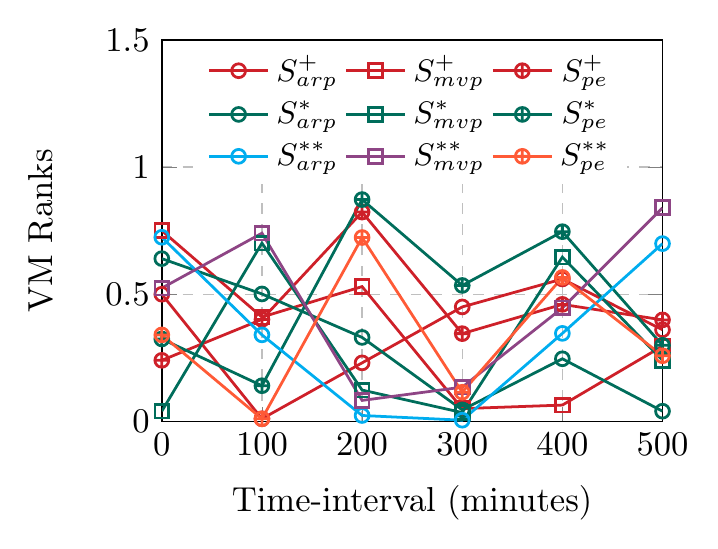}} 
	\subfigure[HA Strategies]{\includegraphics[width=0.69\linewidth, scale=2]{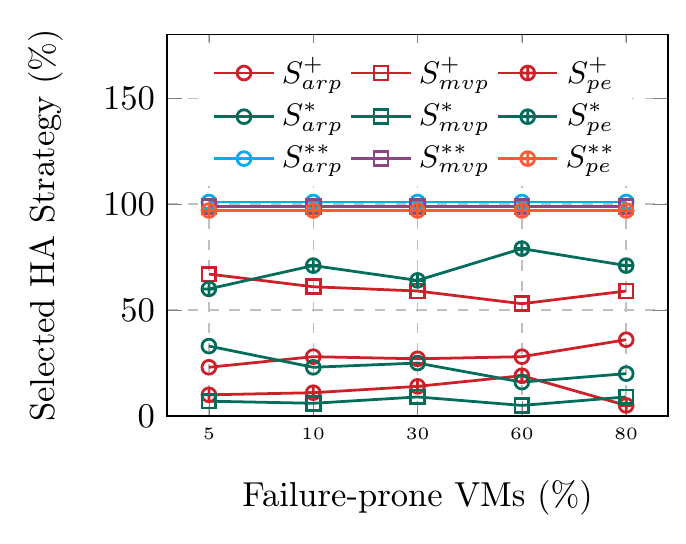}}
	\caption{ HA Strategy Selection}
	\label{rank}		
\end{figure}

Fig. \ref{rank} highlights the comparison of the inter-mediate outcomes achieved during execution of SRE-HM and its various versions: SRE-HM ($S^+$), $S^\ast$, $S^-$, $S^{\ast\ast}_{arp}$, $S^{\ast\ast}_{mvp}$, and $S^{\ast\ast}_{pe}$. Fig. \ref{rank}(a) shows values of  significance ranks obtained for a randomly selected VMs retrieved over a period of 500 minutes via experimental execution. It is observed that significance ranks of a VM varies depending upon the number of invocations  between 0 and 1 over period of 500 minutes. The comparative intermediate resultant values of HA strategy selection (\%)  for varying number of failure-prone VMs (\%) are shown in Fig. \ref{rank}(b), where $S^{\ast\ast}_{arp}$, $S^{\ast\ast}_{mvp}$, $S^{\ast\ast}_{pe}$ show straight lines i.e., 100 \% because these set-ups lack VM rank estimation and HA strategy selection and executed with respect to single HA strategy only.
\subsubsection{HA Analysis}
\begin{figure}[!htbp]
	\centering
	\includegraphics[width=0.69\linewidth]{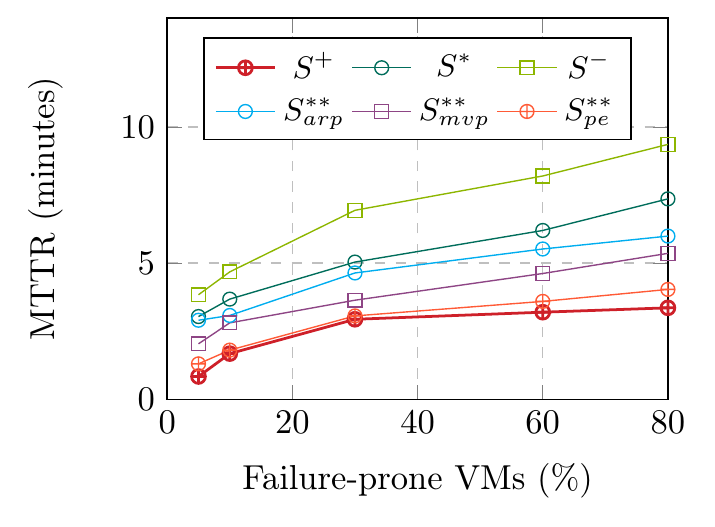}
	\caption{MTTR }
	\label{fig:mttr}
\end{figure}

\begin{figure}[!htbp]
	\centering
	\includegraphics[width=0.69\linewidth]{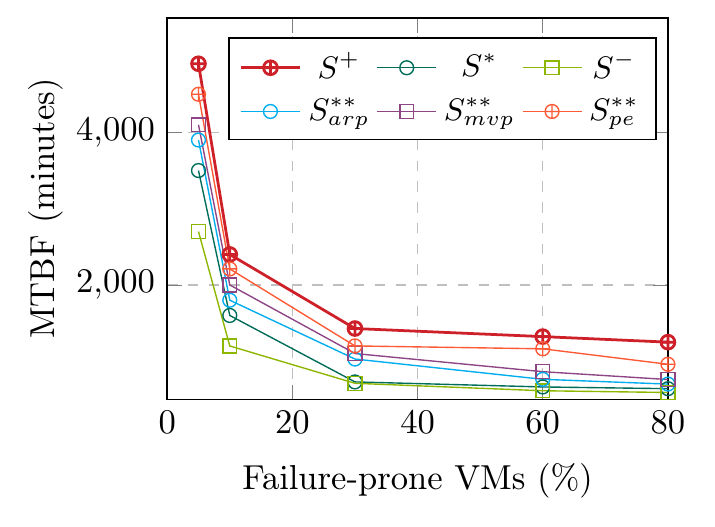}
	\caption{ MTBF }
	\label{fig:mtbf}
\end{figure}

\begin{figure}[!htbp]
	\centering
	\includegraphics[width=0.69\linewidth]{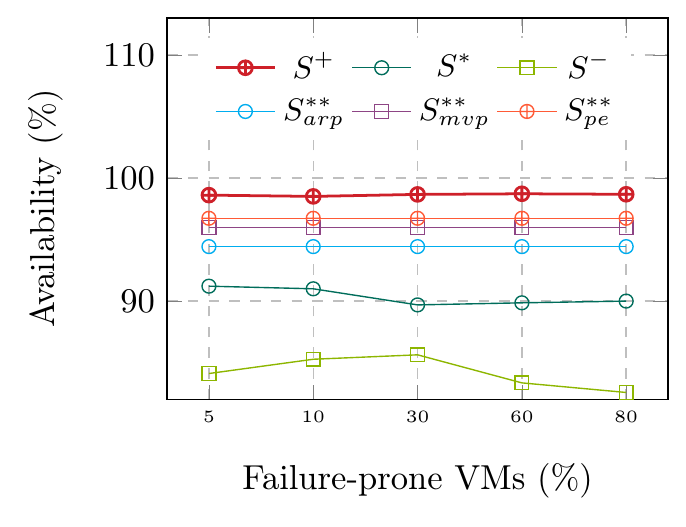}
	\caption{Availability }
	\label{fig:availability}
\end{figure}

The availability varies with the values of MTBF and MTTR, obtained during online processing over time-interval \{$t_1$, $t_2$\}. The variations observed (during experimental simulation) in the values of MTTR and MTBF are shown in Figs. (\ref{fig:mttr}) and (\ref{fig:mtbf}) for Google  Cluster workload execution. It is to be noticed that MTTR decreases when MTBF increases, which specifies a inverse relation between them. The MTBF increases and MTTR decreases in the order: $S^+$>$S^{\ast\ast}_{pe}$ >$S^{\ast\ast}_{mvp}$>$S^{\ast\ast}_{arp}$>$S^\ast$ > $S^-$. Accordingly, 
Fig. \ref{fig:availability}  compares $S^+$, $S^\ast$, $S^{\ast\ast}_{arp}$, $S^{\ast\ast}_{mvp}$, $S^{\ast\ast}_{pe}$, and ($S^-$), where the percentage of availability is highest for $S^+$ and outperforms $S^-$ by 19.56\% due to proactive PM failure estimation and adoption of most appropriate HA strategy on the basis of significance ranking of VMs. 


\subsubsection{Power Consumption}
\begin{figure}[!htbp]
		\centering
	\includegraphics[width=0.69\linewidth]{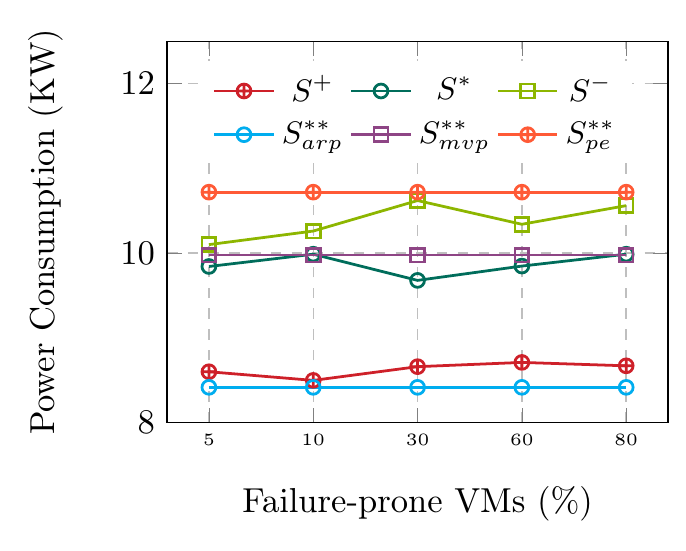}
	\caption{Power Consumption}
	\label{HA analysis}	
\end{figure}

The power consumption is shown in Fig. \ref{HA analysis} which is lesser for $S^+$ up to 19.1\% as compared with that of $S^-$. Further, it is observed that power consumption of $S^+$ is nearly 0.6\% higher as compared with $S^{\ast\ast}_{arp}$ due to adoption of ARP HA strategy for each VM which has lesser power consumption as compared with MVP, and PE strategies. However, at the cost of 0.6\% increased  power consumption, the availability is improved up to 5.31\% by utilizing SRE-HM ($S^+$). 

\subsubsection{Cost Optimization}
\par
\begin{figure}[!htbp]
	\centering	
	\subfigure[Resource Utilization]{\includegraphics[width=0.69\linewidth, scale=2]{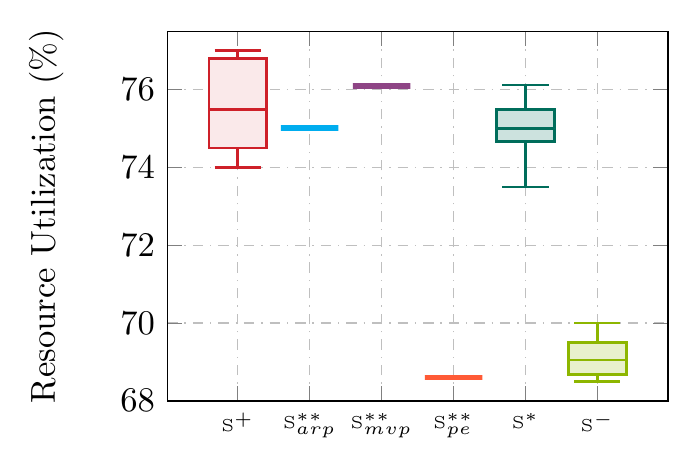}} 
	\subfigure[Active Servers]{\includegraphics[width=0.69\linewidth, scale=2]{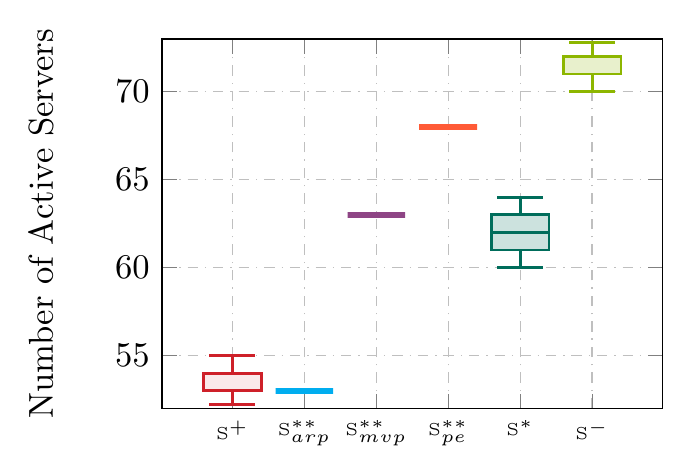}}
	\caption{Cost Optimization parameters}
	\label{cost}
\end{figure}

 Fig. \ref{cost} compares the cost optimization in terms of resource utilization and number of active servers, attained by different versions of SRE-HM viz., With SRE-HM ($S^+$), $S^{\ast\ast}_{arp}$, $S^{\ast\ast}_{mvp}$, $S^{\ast\ast}_{pe}$, with HA Without SRE-HM ($S^-$). Fig. \ref{cost} (a) reveals that $S^+$ has highest RU up to 78.96\% which is superior to $S^{\ast\ast}_{pe}$ by 10\%. The number of active PMs in case of $S^+$ are reduced up to 26.67\% as depicted in Fig. \ref{cost} (b) which is due to consolidation of VMs by including ranking before   HA strategy selection for each VM rather than applying same strategy to all.


\subsubsection{SRE-HM v/s State-of-the-arts}
SRE-HM model is compared with significant state-of-the-art works discussed in Section 2 subject to different key characteristics and performance indicators as shown in Table \ref{summaryComparison}. The description of various characteristics and values of performance metrics are obtained from the published versions of these works while some other values are computed based on the values of related metrics.  The comparison of different performance parameters are reported in subsequent subsections which are based on the values obtained during experimental evaluations of SRE-HM and the compared approaches.  
\begin{table*}[!htbp]
	
	\caption{{Pandect comparison: SRE-HM v/s state-of-the-arts}}
	\label{summaryComparison}
	\scriptsize
	\centering
	\resizebox{1.0\textwidth}{!}{
		\centering
		\begin{tabular}{|p{1.8cm}|| p{1.5cm} |p{1.5cm} |p{1.6cm}||p{1.5cm}|p{1.5cm}| p{1.5cm}| p{1.5cm}|p{1.5cm}|}
			\hline
			
			\multirow{2}{*}	{\textbf{Approaches}}&\multicolumn{3}{c||}{\textbf{Key characteristics}}&\multicolumn{5}{c|}{\textbf{Performance indicators}}
			\\	\cline{2-9} 
			&{\textit{Failure Estimation}}&{\textit{HA Approach}}&{\textit{HA Strategy}}, {\textit{Selection}}& {\textit{Availability }}& {\textit{Resource Utilization }} & {\textit{Energy Consumption }}&{\textit{Active PMs}}&{\textit{VM migration}}\\ \hline \hline

			PEFS \cite{marahatta2020pefs}&Deep NN&  Replication&$\times$&$\times$&70\%-72\%&11500.47& $\times$&$\times$\\ \hline
			{FAEE \cite{sharma2019failure}} & Exp. Smooth. & checkpointing  &$\checkmark$ &82\%&$\times$ &9825.68& $\times$&$\checkmark$\\ \hline	
			HDCC \cite{pinto2016hadoop}&SVM and LR& {$\times$} &{$\times$} &{$\times$} &{$\times$} &{$\times$} & $\times$&$\times$\\ \hline
			CDEF \cite{xu2018improving} &MART-GB&  $\times$ &$\times$ &$\times$ &$\times$ &$\times$ &$\times$ &$\times$\\ \hline
			FESTAL \cite{wang2014festal} &$\times$& Backup &$\checkmark$ &95.08\% &Varying &$\times$ &Varying&$\times$\\ \hline
			FASTER \cite{zhu2016fault} &$\times$& Backup
			&$\checkmark$ &98.5\% &Varying &$\times$ &Varying&$\times$\\ \hline
			OP-MLB \cite{saxena2021op} &DE-NN&{$\times$} &{$\times$} &{$\times$} &{64.6\%} &8801.67kWH& 45\%&$\checkmark$ \\ \hline
			WP-SM	 \cite{dabbagh2018energy} &Wiener Filter&{$\times$} &{$\times$} &{$\times$} &{$\times$} &9001.67 kWH& 51.2\%&$\checkmark$ \\ \hline
			OHD-MUP	 \cite{nguyen2017virtual} &Multiple LR&{$\times$} &{$\times$} &{$\times$} &{$\times$} &7701.67 kWH& 31.2\%&$\checkmark$ \\ \hline
			CHARM \cite{zhang2015charm} &$\times$&  Multi-cloud &$\checkmark$ & Varying &$\times$ &$\times$ &$\times$ &$\times$\\ \hline
			CHASE \cite{jammal2015chase} &$\times$& scheduling &$\checkmark$ &99.1\%  &$\times$&$\times$ &$\times$&$\times$ \\ \hline
			\textbf{SRE-HM} & LSTM-NN&VM Ranking &$\checkmark$  &99.3\% &76.9\% & 8660.7 kWH & 53\% &$\checkmark$  \\ \hline
	\end{tabular}}
\end{table*}

\section{Conclusion and Future Directions}
A novel SRE-HM model is proposed which computes significance ranks of VMs. Based on this ranking,  user specified constraints, and failure probability, an optimal HA strategy is chosen to provide admissible high availability respective to each VM dedicated for execution of cloud applications. A resource predictor detects resource contention prior to its occurrence and alleviates it by migrating VMs from corresponding PM. Implementation and performance evaluation of proposed model acknowledge that it substantially improves the availability for users with optimized operational cost for data centre.  In future, the proposed model can be extended with more useful characteristics of cloud applications to determine their critical nature before siginificance rank computation and VM collaboration relation during resource usage prediction  to achieve more robust high availability at optimized cost.  


%

\appendices

\section*{Acknowledgment}
This work is financially supported by National Institute of Technology, Kurukshetra, India.


\ifCLASSOPTIONcaptionsoff
  \newpage
\fi

\bibliographystyle{IEEEtran}
\bibliography{bibfile}
\begin{IEEEbiography}[{\includegraphics[width=0.7\linewidth]{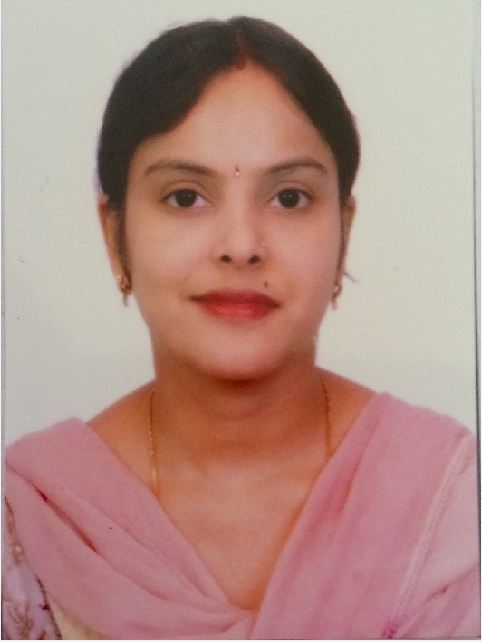}}]{Deepika Saxena}
	received her M.Tech degree in Computer Science and Engineering  from Kurukshetra University Kurukshetra, Haryana, India in 2014. Currently, she is pursuing her Ph.D from Department of Computer Applications, National Institute of Technology (NIT), Kurukshetra, India. Her major research interests are Neural Networks, Evolutionary Algorithms, Resource Management and Security in Cloud Computing.
\end{IEEEbiography}

\begin{IEEEbiography}[{\includegraphics[width=0.7\linewidth]{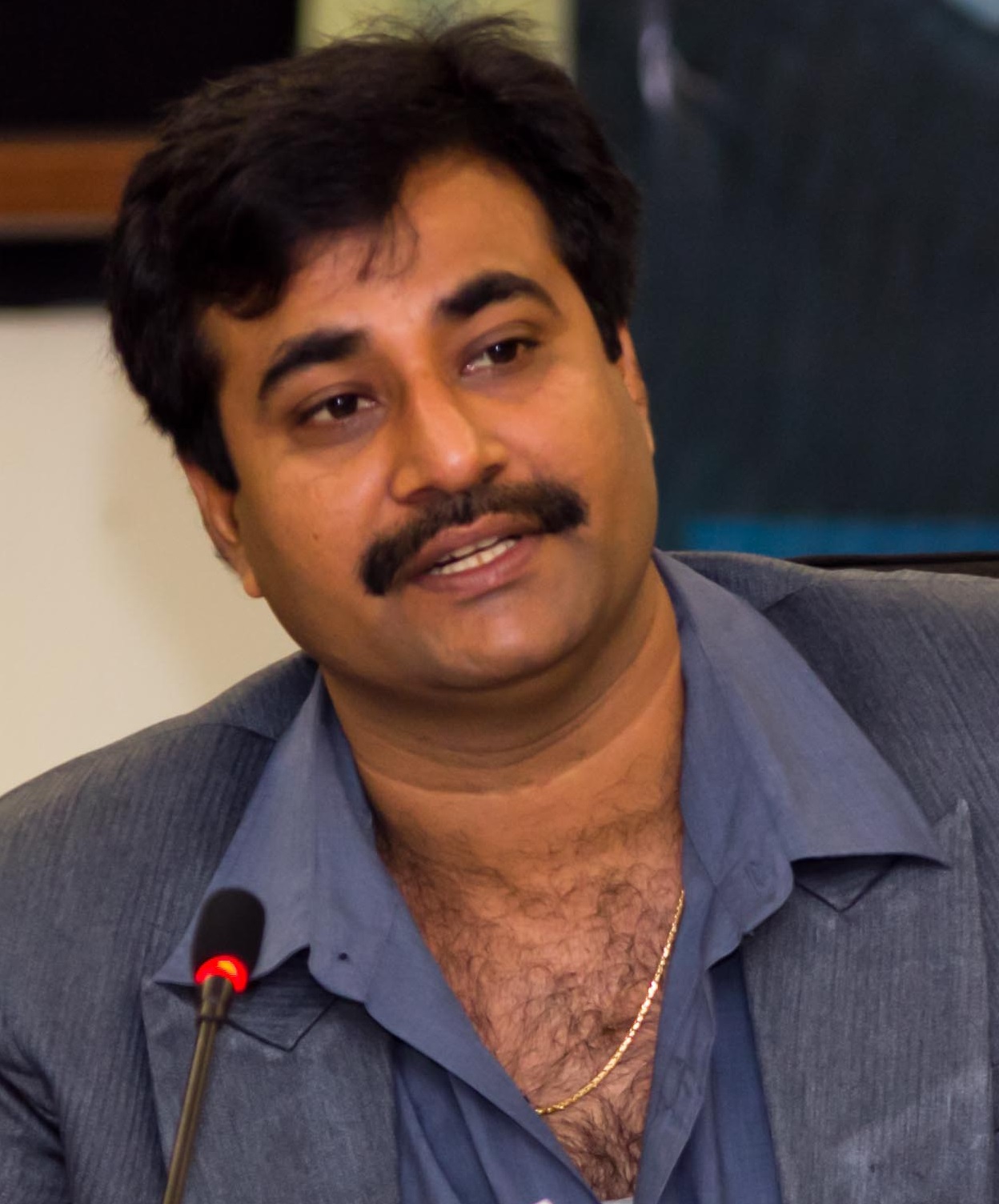}}]{Ashutosh Kumar Singh}
	is working as a Professor and Head in the Department of Computer Applications, National Institute of Technology Kurukshetra, India. He has research and teaching experience in various Universities of the India, UK, and Malaysia. He received his PhD in Electronics Engineering from Indian Institute of Technology, BHU, India and Post Doc from Department of Computer Science, University of Bristol, UK. He is also Charted Engineer from UK. His research area includes Verification, Synthesis, Design and Testing of Digital Circuits, Data Science, Cloud Computing, Machine Learning, Security, Big Data. He has published more than 330 research papers in different journals, conferences and news magazines. 
	
\end{IEEEbiography}
\end{document}